\documentclass[prd,onecolumn,nofootinbib,showpacs]{revtex4}
\usepackage{epsfig,graphicx,bm}
\usepackage[latin2]{inputenc}
\usepackage{color}
\usepackage{t1enc}
\usepackage{amssymb}
\usepackage{amsmath}
\usepackage{graphicx}
\usepackage{bm}
\usepackage{amsfonts}

\newcommand{\bea}{\begin{eqnarray}}
\newcommand{\eea}{\end{eqnarray}}
\newcommand{\be}{\begin{equation}}
\newcommand{\ee}{\end{equation}}
\newcommand{\no}{{\nonumber}}
\newcommand{\eref}[1]{Eq.~\eqref{#1}}
\newcommand{\secref}[1]{Sec.~\ref{#1}}
\newcommand{\apref}[1]{App.~\ref{#1}}
\newcommand{\tabref}[1]{Table~\ref{#1}}
\newcommand{\tr}{\mathop{\mathrm{Tr}}}

\newcommand{\slD}{\mbox{$D$ \hspace{-1.3em} $\slash$}}

\begin{document}

\title{Baryon octet and decuplet phenomenology in a three-flavor
  extended linear sigma model}
\author{$\text{P. Kov{\'a}cs}$}
\email{kovacs.peter@wigner.mta.hu}
\author{$\text{{\'A}. Luk{\'a}cs}$}
\email{lukacs.arpad@wigner.mta.hu}
\author{$\text{J. V{\'a}r{\'o}czy}$}
\email{jannoczy@gmail.com}
\author{$\text{Gy. Wolf}$}
\email{wolf.gyorgy@wigner.mta.hu}
\author{$\text{M. Z{\'e}t{\'e}nyi}$}
\email{zetenyi.miklos@wigner.mta.hu}
\affiliation{Institute for Particle and Nuclear Physics, Wigner Research Center for
Physics, Hungarian Academy of Sciences, H-1525 Budapest, Hungary}

\begin{abstract}
  We present an effective model, which is an extension of the usual
  linear sigma model, that contains a low energy multiplet for every
  hadronic particle type. These multiplets are a scalar nonet, a
  pseudoscalar nonet, a vector nonet, an axialvector nonet, a baryon
  octet and a baryon decuplet. Tree level baryon masses and possible
  two body decuplet decays are calculated. The baryon masses are
  generated through spontaneous symmetry breaking. The calculated
  quantities are used to determine the model parameters through a
  multiparametric minimalization process, which compares the
  calculated physical quantities with their experimental values. We
  found that the calculated quantities are in good agreement with the
  experimental data.
\end{abstract}

\pacs{12.39.Fe, 12.40.Yx, 14.20.Dn, 14.20.Jn}

\maketitle

\allowdisplaybreaks

\section{INTRODUCTION}
\label{Sec:intro}

The phase diagram of QCD, the theory of strong interaction, is a
heavily studied field both theoretically (see
e.g. \cite{Pisarski:1984, Lenaghan:2000, Karsch:2003,
  Schaefer:2004, Barducci:2005, Antoniou:2005, Aoki:2006,
  deForcrand:2006, Szep, Struber:2008} and references therein) and
experimentally (see e.g. \cite{Homma:2007, Toneev:2007, Friese:2012,
  Melkumov:2012, Odyniec:2013} and references therein). Our aim is to develop a
model for that problem, which also reproduces the vacuum phenomenology.

QCD can be solved perturbatively only at very high energies. Although
it is possible to solve QCD nonperturbatively on the lattice, that is
computationally demanding and not very well suited for instance for
scattering problems, or for high densities. We are therefore left with
effective theories. The underlying principle in the construction of
such theories is that they share the same global symmetries as QCD
\cite{Geffen_1969}.

For massless quarks (which is a very good approximation for $u$ and
$d$ and less good for $s$ quarks) the global symmetry of QCD is
$U(N_f)_R \times U(N_f)_L \equiv U(N_f)_V \times U(N_f)_A$, the
so-called chiral symmetry ($R$ stands for right-handed, $L$ for
left-handed quark flavors, and $N_f$ denotes the number of massless
quark flavors). The $U(1)_A$ part of the symmetry is broken by
topological charges \cite{Hooft}, while in the vacuum $SU(3)_A$ is
spontaneously broken \cite{SSB} due to the existence of
quark-antiquark condensates.

There are different ways in which the chiral symmetry of QCD can be
realized. In the QCD Lagrangian, the symmetry is linearly realized,
while in the vacuum and at low energies, the symmetry is nonlinearly
realized. Linear realizations of chiral symmetry have the property
that states are doubled. In nonlinear realizations
\cite{nonlin_sigma}, there can be
states without associated chiral partners. Around the phase transition
the chiral partners are degenerate, so none of them is
negligible. Therefore, in order to investigate the mechanism of chiral
symmetry restoration, which is one of our final aims, effective
theories with linearly realized chiral symmetry \cite{gellmanlevy} are
most appropriate.

The last version of our model \cite{elsm_2013} contained the scalar,
pseudoscalar, vector- and axial-vector nonets of mesons. That model
described the vacuum phenomenology of mesons very well. In this paper
we include the nucleon-octet and the Delta-decuplet to extend the
vacuum phenomenology for baryons as well. Other investigations
concerning on baryon phenomenology can be found for instance in
\cite{Lam:1971, Chemtob:1986, Pirjol:1997, Kalman:1999, Zhuang_1999,
  Hyodo:2005, Riska:2005, Guo:2008, GonzalezdeUrreta:2010, Gallas:2010}.

Our paper is organized as follows. In \secref{Sec:model} we describe
our model with some of the details taken from \cite{elsm_2013}
relegated to \apref{App:meson_lagr}. \secref{Sec:masses_width} is
dedicated to calculation of tree-level baryon masses and decay widths,
while \secref{Sec:results} contains our results of the fitting
procedure. We conclude in \secref{Sec:conclusion}.

\section{THE MODEL}
\label{Sec:model}

The model construction is based on the idea of inclusion of the lowest
lying multiplets for every hadronic particle type, where we assume
that mesons are $q\bar{q}$ and baryons are $qqq$ states. This means
that for mesons we included a scalar, a pseudoscalar, a vector and an
axialvector nonet, while for baryons an octet and a
decuplet. Accordingly, our Lagrangian consist of a mesonic and a
baryonic part, the latter also includes the baryon-meson interaction terms,
$\mathcal{L} = \mathcal{L}_{\text{meson}} +
\mathcal{L}_{\text{baryon}}$, from which we already constructed and
analyzed the meson part in \cite{elsm_2013}, and it is presented
briefly in \apref{App:meson_lagr}.

The Lagrangian of the baryon sector is constructed as follows.
In addition to the kinetic terms of the octet and decuplet baryons we
included such interaction terms with the lowest possible dimension,
that either describe decuplet decays into one octet baryon and one
(pseudo)scalar -- which are the physically relevant two-body decays of
the decuplet --, or such baryon-(pseudo)scalar interactions that
generate octet/decuplet mass terms via spontaneous symmetry breaking
(SSB). The lowest possible dimension for the decuplet decay terms is
four containing one vector-spinor, one spinor, and one (pseudo)scalar
field, and together with the kinetic terms are taken from the leading
order expansion of the nonlinear sigma model \cite{nonlin_sigma} (for
more details see \apref{App:nonlin_expans}). In case of the
baryon-(pseudo)scalar interaction terms (that can produce
octet/decuplet mass terms) the lowest possible dimension is five,
containing two spinor and two (pseudo)scalar fields. Correspondingly,
we included every possible $SU(3)_{V}$ invariants \cite{SU3_inv},
which can be constructed with the given number of fields (see
e.g. App.~C of \cite{Semke_thesis}).

\subsection{Lagrangian}
\label{Subsec:lagrangian}

The baryonic part of the Lagrangian reads

\begin{align}
  \mathcal{L}_{\text{baryon}} & = \tr \left[\bar{B}\left(i \slD-M_{(8)}\right)B\right] \no\\
  & -  \tr\left\{\bar{\Delta}_{\mu}\cdot \left[ \left(i \slD -
        M_{(10)}\right)g^{\mu\nu} - i\left(\gamma^{\mu}D^{\nu} + \gamma^{\nu}D^{\mu}\right)+ 
      \gamma^{\mu}\left(i \slD + M_{(10)}\right)\gamma^{\nu}\right]\Delta_{\nu}\right\} \no\\
  & +  C\tr\left[\bar{\Delta}^{\mu}\cdot\left(
      -\frac{1}{f}(\partial_{\mu} - ieA^{e}_{\mu}[T_3,\Phi]) -
      \frac{1}{f}[\Phi,V_{\mu}] + A_{\mu} \right)B\right] +
  \text{h. c.}  \no \\
  & -  \xi_1^{} \tr \left(\bar{B}B\right ) \tr \left(
    \Phi^{\dagger}\Phi\right) - \xi_2^{} \tr \left( \bar{B} \{\{\Phi,
    \Phi^{\dagger}\}, B \} \right) - \xi_3^{} \tr \left( \bar{B}
    [\{\Phi,\Phi^{\dagger}\},B] \right)\no\\
  & -  \xi_4^{} \left(\tr \left(\bar{B}\Phi\right ) \tr \left(
    \Phi^{\dagger}B\right) + \tr \left(\bar{B}\Phi^{\dagger}\right ) \tr \left(
    \Phi B\right) \right) -  \xi_5^{} \tr \left(\bar{B} \{ [\Phi,
\Phi^{\dagger}], B \} \right) \label{eq:lagr_baryon}\\
  & -  \xi_6^{} \tr \left(\bar{B} [[\Phi,\Phi^{\dagger}], B] \right) -
  \xi_7^{} \left(\tr \left(\bar{B}\Phi\right ) \tr \left( \Phi^{\dagger}B\right) - \tr
    \left(\bar{B}\Phi^{\dagger}\right ) \tr \left( \Phi B\right) \right)\no \\
  & -  \xi_8^{} \left(\tr \left(\bar{B}\Phi B \Phi^{\dagger} \right ) -
    \tr \left(\bar{B}\Phi^{\dagger} B \Phi \right) \right) + \chi_1^{}
    \tr \left( \bar{\Delta} \cdot \Delta \right) \tr \left( \Phi^{\dagger}\Phi\right)\no \\
  & +  \chi_2^{} \tr \left( (\bar{\Delta} \cdot \Delta)
    \{\Phi,\Phi^{\dagger}\} \right) + \chi_3^{} \tr \left( (\bar{\Delta} \cdot
    \Phi) (\Phi^{\dagger} \cdot \Delta) + (\bar{\Delta} \cdot
    \Phi^{\dagger}) (\Phi \cdot \Delta) \right) \no\\
  & +  \chi_4^{} \tr \left( (\bar{\Delta} \cdot \Delta) [\Phi,\Phi^{\dagger}] \right), \no
\end{align}
where $B, \Delta_{\mu}, \Phi, V_{\mu}, A_{\mu}, A^{e}_{\mu}$ stands for the baryon octet, the baryon
decuplet, the scalar-pseudoscalar meson octet, the vectormeson octet, the
axialvector meson octet and the electromagnetic field,
respectively. $M_{(8)}$, $M_{(10)}$ are the bare baryon octet and
decuplet masses, $f$ is related to the pion decay constant, while
$T_a$ denotes the $SU(3)$ generators, $[\; ,\;]$ and $\{\; ,\; \}$
denote commutator and anticommutator, respectively. Moreover, the
covariant derivatives are defined as 
\begin{align}
  D_{\mu} B & = \partial_{\mu} B + i [B,V_{\mu}] + \frac{1}{f}\left\{
    [A_{\mu},\Phi],B\right\}, \no \\
  D_{\mu} \Delta_{\nu}^{ijk}& = \partial \Delta_{\nu}^{ijk} +
  \left(\frac{1}{f} [A_{\mu},\Phi]^{i}_{l} - iV_{\mu\,l}^{\
      \,i}\right) \Delta_{\nu}^{ljk} + \left(\frac{1}{f}
    [A_{\mu},\Phi]^{j}_{l} - iV_{\mu\,l}^{\ \,j}\right)
  \Delta_{\nu}^{ilk} + \left(\frac{1}{f} [A_{\mu},\Phi]^{k}_{l} -
    iV_{\mu\,l}^{\ \,k}\right) \Delta_{\nu}^{ijl}, \no 
\end{align}
and we used the following dot notation:
\be
  (\bar{\Delta} \cdot \Delta)_{k}^{m} \equiv
  \bar{\Delta}_{ijk}\Delta^{ijm}, \quad (\bar{\Delta} \cdot
  \Phi)_{k}^{m} \equiv \bar{\Delta}_{ijk} \Phi_{l}^{i}\epsilon^{jlm},
  \quad (\Phi \cdot \Delta)_{k}^{m} \equiv \Delta^{ijm}
  \Phi_{i}^{l}\epsilon_{jlm}. 
\ee
The explicit forms of the baryon multiplets are as follows
\be
B=\sqrt{2}\sum_{i=1}^{8}B_{a}T_{a}=\left(
\begin{array}{ccc}
\ \frac{1}{\sqrt{2}}\Sigma^{0}+\frac{1}{\sqrt{6}}\Lambda^{0} & \Sigma^{+} & p \\
\ \Sigma^{-} &  -\frac{1}{\sqrt{2}}\Sigma^{0}+\frac{1}{\sqrt{6}}\Lambda^{0} & n \\
\ -\Xi^{-} & \Xi^{0} & -\frac{2}{\sqrt{2}}\Lambda^{0}
\end{array}
\right),
\ee
\bea
\Delta^{111}_{\mu} &=& \Delta^{++}_{\mu},\quad
\Delta^{112}_{\mu} = \frac{1}{\sqrt{3}}\Delta^{+}_{\mu}, \quad \Delta^{122}_{\mu} =
\frac{1}{\sqrt{3}}\Delta^{0}_{\mu}, \quad \Delta^{222}_{\mu} =
\Delta^{-}_{\mu},\\
\Delta^{113}_{\mu} &=& \frac{1}{\sqrt{3}}\Sigma^{\star+}_{\mu}, \quad
\Delta^{123}_{\mu} = \frac{1}{\sqrt{6}}\Sigma^{\star0}_{\mu}, \quad
\Delta^{223}_{\mu} = \frac{1}{\sqrt{3}}\Sigma^{\star-}_{\mu},\\
\Delta^{133}_{\mu} &=& \frac{1}{\sqrt{3}}\Xi^{\star0}_{\mu}, \quad
\Delta^{233}_{\mu} = \frac{1}{\sqrt{3}}\Xi^{\star-}_{\mu},\\ 
\Delta^{333}_{\mu} &=&\Omega^{-}_{\mu}, 
\eea
while the explicit form of the scalar-pseudoscalar octet is
\be
\Phi \equiv \Phi_S + \Phi_P =\sum_{i=0}^{8}(S_{i}+iP_{i})T_{i}=\frac{1}{\sqrt{2}}\left(
\begin{array}
[c]{ccc}%
\frac{(\sigma_{N}+a_{0}^{0})+i(\eta_{N}+\pi^{0})}{\sqrt{2}} & a_{0}^{+}
+i\pi^{+} & K_{0}^{\star+}+iK^{+}\\
a_{0}^{-}+i\pi^{-} & \frac{(\sigma_{N}-a_{0}^{0})+i(\eta_{N}-\pi^{0})}
{\sqrt{2}} & K_{0}^{\star0}+iK^{0}\\
K_{0}^{\star-}+iK^{-} & {\bar{K}_{0}^{\star0}}+i{\bar{K}^{0}} & \sigma
_{S}+i\eta_{S}%
\end{array}
\right), 
\ee
and the remaining two multiplets can be found in \apref{App:meson_lagr}.

An important point here is that in the physical scalar sector of low energy QCD
beside the scalar $q\bar{q}$ octet included in our model there are
other states like glueballs and tetraquarks having similar or even lower mass 
than the $q\bar{q}$ states, which in principle can mix
with their corresponding $q\bar{q}$ partner. However the scalar
tetraquarks, like $f_0(500)$ have a much smaller mass then the diquark
state with the same quantum number, thus we expect that their mixings
are small. The glueball $f_0$ - which should have mass around
$1.5$~GeV - probably has a considerable mixing with the $f_0$ states
considered here (see the discussion in \cite{elsm_2013}), which should
be investigated, but the properties of the scalar sector are not
included in the fitting procedure and are beyond the scope of this
paper. On the other hand, the scalars have no direct influence on the
properties of the baryons considered here, because none of these
baryons has large partial decay widths into scalars and baryons \cite{PDG},
which means either their couplings are weak or the considered scalar
mass is too large. In both cases the contribution of the scalars to
the self-energies of the baryons are small. Thus at first glance a more
precise description of the scalar sector is not essential in this
discussion.  
 
It is worth noted that the pseudoscalar ($P_a$), axialvector
($A_{a}^{\mu}$) and baryon octet ($B_a$) fields are not physically
observable in their current form, since for
example $P_1$ is not observable, only the combination $(P_1 -
iP_2)/\sqrt{2}$, which is $\pi^{+}$. Thus for later calculation it is
worth to transform the above fields into physically observable forms
\footnote{In the $0-8$ sector of the (pseudo)scalars, where there is
  particle mixing, another orthogonal transformation is needed in
  order to transform them into physically observable particles}, as already shown in
their matrix form. This can be
done with the following $8\times 8$ (in case of the baryon octet) and
$9\times 9$ (in case of the meson nonets) transformations as
\begin{align}
 Q^{(8)} &= \mathrm{diag}\left( \frac{1}{\sqrt{2}} \begin{pmatrix}
     1&-i \\ 1&i \end{pmatrix}, 1,
   \frac{1}{\sqrt{2}} \begin{pmatrix} 1&-i \\ 1&i \end{pmatrix},
   \frac{1}{\sqrt{2}} \begin{pmatrix} 1&-i \\ 1&i \end{pmatrix}, 1
   \right),\no \\
 Q^{(9)} &= \mathrm{diag}\left(1, \frac{1}{\sqrt{2}} \begin{pmatrix}
     1&-i \\ 1&i \end{pmatrix}, 1,
   \frac{1}{\sqrt{2}} \begin{pmatrix} 1&-i \\ 1&i \end{pmatrix},
   \frac{1}{\sqrt{2}} \begin{pmatrix} 1&-i \\ 1&i \end{pmatrix}, 1
   \right),\label{eq:Q_transf} 
\end{align}
with which the fields can be written as
\begin{align}
  P_{A} &= Q^{(9)}_{Aa} P_a = \left( P_0, \pi^{+}, \pi^{-}, \pi^{0},
    K^{+}, K^{-}, K^{0}, \bar{K}^{0}, P_8  \right),\no \\
  A_{A}^{\mu} &= Q^{(9)}_{Aa} A_a^{\mu} = \left( A_0, a_1^{+}, a_1^{-}, a_1^{0},
    K_1^{+}, K_1^{-}, K_1^{0}, \bar{K_1}^{0}, A_8  \right)^{\mu},\\
  B_{A} &= Q^{(8)}_{Aa} B_a = \left( \Sigma^{+}, \Sigma^{-}, \Sigma^{0},
    p, -\Xi^{-}, n, \Xi^{0}, \Lambda^8  \right).\no
\end{align}

As one can see from the Lagrangian, the baryonic sector has 16 -- yet
unknown -- parameters: two bare masses $M_{(8)}$ and $M_{(10)}$; eight
octet baryon-(pseudo)scalar couplings $\xi_1,\ldots, \xi_8$; four
decuplet baryon-(pseudo)scalar couplings $\chi_1,\ldots ,\chi_4$; one
$\Delta$-decay constant $C$, and the parameter $f$. However, as one
shall see in the next chapter not all the 16 parameters are
independent or some of them not even appear in the formulas of the
physical quantities considered here.  

\section{Baryon masses and decuplet decays}
\label{Sec:masses_width}

After the Lagrangian is fixed, as a usual procedure we require non-zero
vacuum expectation values for certain scalar fields, namely for the
non strange $\sigma_N$ and strange $\sigma_S$ scalar
fields\footnote{We use the so-called non strange--strange basis defined
in \eref{eq:nsbase}.}, which
corresponds to the isospin symmetric case (see
e.g. \cite{Kovacs_2007}). The vacuum expectation values will be
denoted by
\be
 \phi_N\equiv <\sigma_N>,\quad \phi_S\equiv<\sigma_S>. \label{eq:vev}
\ee 
Than one should shift the $\sigma_N$, $\sigma_S$ scalar fields by
their expectation values in the Lagrangian in order to get the
tree-level masses and decay widths around the true vacuum,
\be
 \sigma_N \to \sigma_N + \phi_N,\quad \sigma_S \to \sigma_S + \phi_S\label{eq:sigma_shift}.
\ee 
It is easy to see that the terms proportional to $\xi_5, \xi_6, \chi_4$
and $\xi_7, \xi_8$ do not contribute to the masses. In case of the
first three terms it is due to the fact that $[T_0,T_8]=0$, while in
case of the second two terms it is due to the scalar octet is
hermitian ($\Phi_S=\Phi_S^{\dagger}$). Moreover, in the expression of
the baryon octet masses $\xi_1$ and $M_{(8)}$, while in case of
the decuplet baryon masses $\chi_1$ and $M_{(10)}$ always appear
in the same combination, thus without loss of generality we can
set $\xi_1=\chi_1=0$. Although, in scattering processes both $\xi_1$
and $\chi_1$ would be needed, but these processes are not considered here. 

After some straightforward calculation the terms quadratic in the
fields $B$ and $\Delta_{\mu}$ can be determined, and consequently the
three-level baryon masses are found to be
\begin{align}
  m_p = m_n &= M_{(8)} + \frac{1}{2}\xi_2(\Phi_N^2 + 2\Phi_S^2) +
  \frac{1}{2}\xi_3 (\Phi_N^2 - 2\Phi_S^2), \\
  m_{\Xi} &= M_{(8)} + \frac{1}{2}\xi_2(\Phi_N^2 + 2\Phi_S^2) -
  \frac{1}{2}\xi_3 (\Phi_N^2 - 2\Phi_S^2), \\
  m_{\Sigma} &= M_{(8)} + \xi_2 \Phi_N^2, \\
  m_{\Lambda} &= M_{(8)} + \frac{1}{3}\xi_2(\Phi_N^2 + 4\Phi_S^2) +
  \frac{1}{3}\xi_4 (\Phi_N - \sqrt{2}\Phi_S)^2,
\end{align}

\begin{align}
  m_{\Delta} &= M_{(10)}  + \frac{1}{2} \chi_2^{} \Phi_N^2, \\
  m_{\Sigma^{\star}} &= M_{(10)} + \frac{1}{3}\chi_{2}^{} (\Phi_N^2 +
  \Phi_S^2) + \frac{1}{6} \chi_{3}^{} (\Phi_N - \sqrt{2}\Phi_S)^2, \\
  m_{\Xi^{\star}} &= M_{(10)} + \frac{1}{6} \chi_{2}^{} (\Phi_N^2 + 4
  \Phi_S^2) + \frac{1}{6} \chi_{3}^{} (\Phi_N - \sqrt{2}\Phi_S)^2, \\
  m_{\Omega} &= M_{(10)} + \chi_{2}^{} \Phi_S^2.
\end{align}

\subsection{Decay widths}
\label{Subsec:decay_width}

According to the PDG \cite{PDG}, one can consider four physically allowed
two-body decays of the decuplet baryons. These are the following
\begin{equation}
\Delta \to p \pi, \quad \Sigma^{\star} \to \Lambda \pi,\quad
\Xi^{\star} \to \Xi \pi, \quad \Sigma^{\star} \to \Sigma
\pi. \label{eq:Delta_decay}
\end{equation}
After applying the field shifts \eref{eq:sigma_shift} in the $C$-term
of the Lagrangian \eref{eq:lagr_baryon} the corresponding interaction
part is given by 
\be 
\mathcal{L}_{\Delta\to B P} = -\frac{C}{f}
G_{ijk}^{ab} \bar{\Delta}_{ijk}^{\mu} (\partial_{\mu}P_{a}) B_{b} +
C \cdot G_{ijk}^{ab} \bar{\Delta}_{ijk}^{\mu} A_{a\,\mu}
B_b,\label{eq:decay_lagr} 
\ee 
where the $G_{ijk}^{ab}$ coupling constant reads as 
\be 
G_{ijk}^{ab} \equiv \frac{\sqrt{2}}{4}
\epsilon_{ilm} (\lambda^a)_{jl} (\lambda^b)_{km}.  
\ee
Looking at \eref{eq:decay_lagr} one could ask why the second term
present, since it does not contain pseudoscalars, however all of the
decays in \eref{eq:Delta_decay} does. Actually, due to a mixing between the
(axial)vectors and the (pseudo)scalars in the meson sector a
redefinition of certain (axial)vector fields is needed (see
\apref{App:meson_lagr} for details), which will bring in the
(pseudo)scalars into the second term (see \eref{eq:a1_shift}). 

Using the \eref{eq:Q_transf} transformations and the field
redefinitions \eref{eq:a1_shift} in \eref{eq:decay_lagr} the
resulting Lagrangian is
\begin{equation}
  \begin{split} 
    \mathcal{L}_{\Delta\to B P} &=
    \frac{G}{\sqrt{2}} \Delta_{\mu}^{--} (\partial^{\mu}\pi^{+}) p -
    \frac{G}{2}\Sigma_{\mu}^{\star\, -} (\partial^{\mu}\pi^{+}) \Lambda^0
     \\ &- \frac{G}{2\sqrt{3}}\Sigma_{\mu}^{\star\, -} \left[
      (\partial^{\mu}\pi^{+}) \Sigma^0 + (\partial^{\mu}\pi^{0}) \Sigma^{+}
    \right ] + \frac{G}{2\sqrt{3}} \Xi_{\mu}^{\star\, -} \left[
      (\partial^{\mu}\pi^{+}) \Xi^{-} + (\partial^{\mu}\pi^{0}) \Xi^{0}
    \right ],\label{eq:lagr_decay_fin}
  \end{split}
\end{equation}
with
\be
G = CZ_{\pi}\left(g_1 w_{a_1}-\frac{i}{f} \right),\no
\ee 
where $Z_{\pi}$ and $w_{a_1}$ are defined in
\apref{App:meson_lagr}. Moreover, terms including the same decaying
particle with different charges are not written out, since they would result in the
same decay widths. According to \eref{eq:form_decay_width} the decay
width can be calculated as
\be
\Gamma_{\Delta \to PB} = I_{\Delta \to PB} \frac{k^3}{12 m_{\Delta}} (m_{B} + E_{B})
|G_{(\Delta \to PB)}|^2,
\ee
where $k$ and $E_{B}$ are given by \eref{eq:k} and \eref{eq:E}, while the
isospin factor $I_{\Delta \to PB}$ is one for $\Delta^{++} \to p
\pi^+$ and $\Sigma^{\star\, +} \to \Lambda \pi^+$, two for
$\Sigma^{\star +} \to \Sigma^{\underset{+}{0}} 
\pi^{\overset{+}{0}}$, since there are two channels
$\Sigma^+ \pi^0$ and $\Sigma^0 \pi^+$, and three for $\Xi^{\star\, 0}
\to \Xi^{\overset{-}{0}} \pi^{\overset{+}{0}}$, where there is
one charged $\Xi^- \pi^+$ and one neutral channel $\Xi^0
\pi^0$. Accordingly, for the decays of \eref{eq:Delta_decay}, the decay
widths are given by
\begin{align}
  \Gamma_{\Delta \to \pi p} &= \frac{k^3_{p}}{24 m_{\Delta}}
  (m_{p} + E_{p}) G^2,\quad 
  \Gamma_{\Sigma^{\star} \to \pi \Lambda} =
  \frac{k^3_{\Lambda}}{48 m_{\Sigma^{\star}}} (m_{\Lambda} +
  E_{\Lambda})  G^2,\no \\
  \Gamma_{\Xi^{\star} \to \pi \Xi} &= \frac{k^3_{\Xi}}{48
    m_{\Xi^{\star}}} (m_{\Xi} + E_{\Xi}) G^2,\quad
  \Gamma_{\Sigma^{\star} \to \pi \Sigma} =
  \frac{k^3_{\Sigma}}{72 m_{\Sigma^{\star}}} (m_{\Sigma} +
  E_{\Sigma}) G^2,
\end{align}
with
\be
G^2 = C^2Z_{\pi}^2\left(g_1^2 w_{a_1}^2+\frac{1}{f^2} \right).\no
\ee

\section{$\chi^2$-fit and results}
\label{Sec:results}

In the fitting procedure we used a $\chi^2$ minimalization method to
determine the parameters of the baryon Lagrangian as we did for the
meson Lagrangian in \cite{elsm_2013} from where we took the parameters
of the mesonic sector. Our aim was to find a parameter
set with which the calculated values of the observables deviate from
their experimental values only within a given error. Since, isospin
breaking is neglected, our calculation is at tree-level and our
model is an effective model of QCD, not the QCD itself, we do not
expect that it reproduces all the observables perfectly. Accordingly,
we artificially set the errors to 5\% for the masses and to 10\% for the
decay width, since they have a larger uncertainty.   

In the baryon Lagrangian there are 8 unknown parameters, namely,
$M_{(8)}$, $\xi_2$, $\xi_3$ and  $\xi_4$ are describing the octet masses,
$M_{(10)}$, $\chi_2$ and $\chi_3$ the decuplet masses, while $G\equiv
CZ_{\pi}\sqrt{g_1^2 w_{a_1}^2+1/f^2}$ the decay widths. In order to
determine these parameters we define the $\chi^2$ as
\begin{equation}
  \chi^2(x_1,\dots,x_N)=\sum_{i=1}^{M}\left[\frac{Q_i(x_1,\dots,x_N)-Q_i^{\text{exp}}}{\delta
      Q_i}\right]^2,
\end{equation}
where $x_1,\dots,x_N$ are the unknown parameters, the $M$ observables
$Q_i(x_1,\dots,x_N)$ are calculated from the model, while
$Q_i^{\text{exp}}$ are taken from the PDG \cite{PDG} with the chosen
error $\delta Q_i$ as discussed above. For the multiparametric
minimalization of $\chi^2(x_1,\dots,x_N)$ the MINUIT \cite{MINUIT} code was
used. In this particular case we have 8 parameters to fit for the 12
observables. The resulting parameters are given in
\tabref{Tab:param_baryon} along with their theoretical errors, which
characterize how sensitive quantities are to the change of the
given variable. For instance the large error of $\chi_3$ in
\tabref{Tab:param_baryon} means that $\chi_3$ should be changed
by $4387.18$ in order to change $\chi^2$ by one. The values of the
observables along with their experimental values and errors can be
found in \tabref{Tab:observables}. 
\begin{table}[th]
\centering
\begin{tabular}
[c]{|c|c|}\hline
Parameter & Value\\\hline
$M_{(8)}\; [\text{GeV}]$ & $1.92\pm 0.05$ \\\hline 
$\xi_2\; [\text{GeV}^{-1}]$ & $-27.01\pm 1.57$ \\\hline
$\xi_3\; [\text{GeV}^{-1}]$ & $79.35\pm 16.70$ \\\hline
$\xi_4\; [\text{GeV}^{-1}]$ & $139.33\pm 1063.42$ \\\hline
$M_{(10)}\; [\text{GeV}]$ &  $-1.27\pm 0.03$ \\\hline 
$\chi_2\; [\text{GeV}^{-1}]$ & $184.42\pm 2.13$ \\\hline
$\chi_3\; [\text{GeV}^{-1}]$ & $213.00\pm 4387.18$ \\\hline
$G\; [\text{GeV}^{-1}]$ & $9.88\pm 2.16$ \\\hline 
\end{tabular}
\caption{Baryon parameters and their theoretical errors}    
\label{Tab:param_baryon}
\end{table}                    
It is important to note that all the parameters appeared already in the
meson sector are fixed during the fit and their values are presented
in \tabref{Tab:param_meson}. 
\begin{table}[th]
\[
\begin{array}[c]{|c|c|c|c|}
\hline 
\text{Observable} & \text{Fit [MeV]}  & \text{Experiment [MeV]} \\\hline
m_p & 939.0\pm 59.6 & 939.0\pm 47.0 \\ \hline
m_{\Lambda} & 1116.0\pm 67.0 & 1116.0\pm 55.8 \\ \hline
m_{\Sigma} & 1193.0\pm 69.3 & 1193.0\pm 59.7 \\ \hline
m_{\Xi} & 1318.0\pm 75.3 & 1318.0\pm 65.9 \\ \hline
m_{\Delta} & 1231.9\pm 58.5 & 1232.0\pm 61.6 \\ \hline
m_{\Sigma^{\star}} & 1385.5\pm 50.6 & 1385.0\pm 69.3 \\ \hline
m_{\Xi^{\star}} & 1532.3\pm 51.1  & 1533.0\pm 76.7 \\ \hline
m_{\Omega} & 1672.3\pm 78.3 & 1672.0\pm 83.6 \\ \hline
\Gamma_{\Delta \rightarrow p \pi} & 72.4\pm 3.5 & 110.0\pm 11.0\\ \hline
\Gamma_{\Sigma^{\star} \rightarrow \Lambda \pi} & 29.1\pm 1.4 & 32.0\pm 3.2 \\ \hline
\Gamma_{\Sigma^{\star} \rightarrow \Sigma \pi} & 3.9\pm 0.2 & 4.3\pm 0.4 \\ \hline
\Gamma_{\Xi^{\star} \rightarrow \Xi \pi} & 12.0\pm 0.6 & 9.5\pm 1.0 \\
\hline
\end{array}
\]
\caption{Calculated and experimental values of the observables along
  with their theoretical and experimental errors}
\label{Tab:observables}
\end{table}
It can be seen from \tabref{Tab:observables} that the octet masses can
be described perfectly, which is not so surprising, since we have four
parameters to fit for four quantity and all the equations are linear
in the parameters. It is more interesting that the decuplet masses are
given with almost the same precision as the octet masses, even if we
have only three independent parameters in this sector to fit. Finally,
for the decuplet decays we have only one parameter for four physical
observables and as expected the tree level expressions, which differ
from each other only in their kinematic parts, can not give
back all the experimental values with a very good precision. The
unnatural values of $M_{(8)}$ and $M_{(10)}$ do not concern us,
since with appropriately chosen values of $\xi_1$ and $\chi_1$ we can achieve any
values for $M_{(8)}$ and $M_{(10)}$.

\section{CONCLUSIONS}
\label{Sec:conclusion}

We have presented a possible baryon octet and decuplet extension to
our previous meson model \cite{elsm_2013}. We included interaction
terms, such as $\Delta-B-P$ suggested by the lowest order chiral
perturbation theory, other interaction terms like the $B-B-\Phi-\Phi$ kind
of terms was introduced in order to generate baryon masses. In the
last case we included every possible $SU(3)_{V}$ invariants. 

From the constructed Lagrangian we calculated the tree-level masses
and physically relevant decuplet decay widths and we found that in
general they are in good agreement with the experimental data taken
from the PDG \cite{PDG}.

As a continuation other interaction terms, which contain derivatives
could also be introduced (see e.g. \cite{Zhuang_1999}), which are
important if one would like to investigate scattering processes as
well. Another interesting direction is to move on to finite
temperature and/or densities with these fields included in our
model. However, this task seems not an easy one. For instance it is
not obvious how one can switch from the baryon octet and decuplet
degrees of freedom, which are the appropriate degrees of freedom at
low temperature and densities, to the constituent quarks, which are
better candidates for degrees of freedom as one approaches the phase
transition region.

\section*{Acknowledgements}

P.\ Kov{\'a}cs, Gy.\ Wolf and M. Z{\'e}t{\'e}nyi were supported by the Hungarian OTKA
funds NK101438 and K109462, while {\'A}. Luk{\'a}cs by OTKA fund K101709.

\appendix

\section{Meson Lagrangian}
\label{App:meson_lagr}

The meson Lagrangian is basically the same, as in \cite{elsm_2013}
with the exception that the dilaton field is completely neglected and
it reads as 
\begin{align}
\mathcal{L_{\text{meson}}}  &
= \tr [(D_{\mu}\Phi)^{\dagger}(D_{\mu}\Phi)]-m_{0}
^{2}\tr(\Phi^{\dagger}\Phi)-\lambda_{1}
[\tr(\Phi^{\dagger}\Phi)]^{2}-\lambda_{2}
\tr(\Phi^{\dagger}\Phi)^{2}{\nonumber}\\
&  -\frac{1}{4}\tr(L_{\mu\nu}^{2}+R_{\mu\nu}^{2}
)+\tr\left[  \left(  \frac{m_{1}^{2}}{2}+\Delta\right)
(L_{\mu}^{2}+R_{\mu}^{2})\right]
+\tr[H(\Phi+\Phi^{\dagger})]{\nonumber}\\
&  +c_{1}(\det\Phi-\det\Phi^{\dagger})^{2}+i\frac{g_{2}}{2}
(\tr\{L_{\mu\nu}[L^{\mu},L^{\nu}
]\}+\tr\{R_{\mu\nu}[R^{\mu},R^{\nu}]\}){\nonumber}\\
&  +\frac{h_{1}}{2}\tr(\Phi^{\dagger}\Phi
)\tr(L_{\mu}^{2}+R_{\mu}^{2})+h_{2}
\tr[(L_{\mu}\Phi)^{2}+(\Phi R_{\mu} )^{2}]+2h_{3}
\tr(L_{\mu}\Phi R^{\mu}\Phi^{\dagger}).{\nonumber}\\
&  +g_{3}[\tr(L_{\mu}L_{\nu}L^{\mu}L^{\nu}
)+\tr(R_{\mu}R_{\nu}R^{\mu}R^{\nu})]+g_{4}
[\tr\left(  L_{\mu}L^{\mu}L_{\nu}L^{\nu}\right)
+\tr\left(  R_{\mu}R^{\mu}R_{\nu}R^{\nu}\right)
]{\nonumber}\\
&  +g_{5}\tr\left(  L_{\mu}L^{\mu}\right)
\,\tr\left(  R_{\nu}R^{\nu}\right)  +g_{6}
[\tr(L_{\mu}L^{\mu})\,\tr(L_{\nu}L^{\nu
})+\tr(R_{\mu}R^{\mu})\,\tr(R_{\nu}R^{\nu
})]\text{  ,} \label{eq:Lagrangian}
\end{align}
where
\begin{align}
D^{\mu}\Phi &  \equiv\partial^{\mu}\Phi-ig_{1}(L^{\mu}\Phi-\Phi R^{\mu
})-ieA^{e\,\mu}[T_{3},\Phi]\;,\nonumber\\
L^{\mu\nu}  &  \equiv\partial^{\mu}L^{\nu}-ieA^{e\,\mu}[T_{3},L^{\nu}]-\left\{
\partial^{\nu}L^{\mu}-ieA^{e\,\nu}[T_{3},L^{\mu}]\right\}\;  ,\nonumber\\
R^{\mu\nu}  &  \equiv\partial^{\mu}R^{\nu}-ieA^{e\,\mu}[T_{3},R^{\nu}]-\left\{
\partial^{\nu}R^{\mu}-ieA^{e\,\nu}[T_{3},R^{\mu}]\right\}
\; ,\nonumber
\end{align}
The quantities $\Phi$, $R^{\mu}$, and $L^{\mu}$ represent the scalar-pseudoscalar,
the left- and right-handed vector nonets:
\begin{align}
\Phi &  =\sum_{i=0}^{8}(S_{i}+iP_{i})T_{i}=\frac{1}{\sqrt{2}}\left(
\begin{array}
[c]{ccc}
\frac{(\sigma_{N}+a_{0}^{0})+i(\eta_{N}+\pi^{0})}{\sqrt{2}} & a_{0}^{+}
+i\pi^{+} & K_{0}^{\star+}+iK^{+}\\
a_{0}^{-}+i\pi^{-} & \frac{(\sigma_{N}-a_{0}^{0})+i(\eta_{N}-\pi^{0})}
{\sqrt{2}} & K_{0}^{\star0}+iK^{0}\\
K_{0}^{\star-}+iK^{-} & {\bar{K}_{0}^{\star0}}+i{\bar{K}^{0}} & \sigma
_{S}+i\eta_{S}
\end{array}
\right)  ,\label{eq:matrix_field_Phi}\\
L^{\mu}  &  =\sum_{i=0}^{8}(V_{i}^{\mu}+A_{i}^{\mu})T_{i}=\frac{1}{\sqrt{2}
}\left(
\begin{array}
[c]{ccc}
\frac{\omega_{N}+\rho^{0}}{\sqrt{2}}+\frac{f_{1N}+a_{1}^{0}}{\sqrt{2}} &
\rho^{+}+a_{1}^{+} & K^{\star+}+K_{1}^{+}\\
\rho^{-}+a_{1}^{-} & \frac{\omega_{N}-\rho^{0}}{\sqrt{2}}+\frac{f_{1N}
-a_{1}^{0}}{\sqrt{2}} & K^{\star0}+K_{1}^{0}\\
K^{\star-}+K_{1}^{-} & {\bar{K}}^{\star0}+{\bar{K}}_{1}^{0} & \omega
_{S}+f_{1S}
\end{array}
\right)  ^{\mu},\label{eq:matrix_field_L}\\
R^{\mu}  &  =\sum_{i=0}^{8}(V_{i}^{\mu}-A_{i}^{\mu})T_{i}=\frac{1}{\sqrt{2}
}\left(
\begin{array}
[c]{ccc}
\frac{\omega_{N}+\rho^{0}}{\sqrt{2}}-\frac{f_{1N}+a_{1}^{0}}{\sqrt{2}} &
\rho^{+}-a_{1}^{+} & K^{\star+}-K_{1}^{+}\\
\rho^{-}-a_{1}^{-} & \frac{\omega_{N}-\rho^{0}}{\sqrt{2}}-\frac{f_{1N}%
-a_{1}^{0}}{\sqrt{2}} & K^{\star0}-K_{1}^{0}\\
K^{\star-}-K_{1}^{-} & {\bar{K}}^{\star0}-{\bar{K}}_{1}^{0} & \omega
_{S}-f_{1S}
\end{array}
\right)  ^{\mu}, \label{eq:matrix_field_R}
\end{align}
where the assignment to physical particles is also shown, except in
the $0-8$ sector, where there is particle mixing \cite{Szep,
  elsm_2013, mixing} and the physical
fields are given by certain orthogonal transformation from the
non-physical fields. Here, $T_{i}\,(i=0,\ldots,8)$ denote the
generators of $U(3)$, while $S_{i}$ represents the scalar, $P_{i}$ the
pseudoscalar, $V_{i}^{\mu}$ the vector, and
$A_{i}^{\mu}$ the axial-vector meson fields, and $A^{e\,\mu}$ is the
electromagnetic field. It should be noted that here and throughout the
article we use the so-called non strange -- strange basis in the
$(0-8)$ sector, defined as
\begin{align}
\varphi_{N}  &  =\frac{1}{\sqrt{3}} \left( \sqrt{2}\; \varphi_{0}+
\varphi_{8}\right)\;,\nonumber\\
\varphi_{S}  &  =\frac{1}{\sqrt{3}} \left(
  \varphi_{0}-\sqrt{2}\;\varphi_{8} \right) \;,\quad\quad
\varphi_{i}\in(S_i,P_i,V_i^{\mu},A_i^{\mu})\;, \label{eq:nsbase}
\end{align}
which is more suitable for our calculations. Moreover, 
$H$ and $\Delta$ are constant external fields defined as
\begin{align}
H  &  =H_{0}T_{0}+H_{8}T_{8}=\left(
\begin{array}
[c]{ccc}
\frac{h_{0N}}{2} & 0 & 0\\
0 & \frac{h_{0N}}{2} & 0\\
0 & 0 & \frac{h_{0S}}{\sqrt{2}}
\end{array}
\right)\;  ,\label{eq:expl_sym_br_epsilon}\\
\Delta &  =\Delta_{0}T_{0}+\Delta_{8}T_{8}=\left(
\begin{array}
[c]{ccc}
\frac{\tilde{\delta}_{N}}{2} & 0 & 0\\
0 & \frac{\tilde{\delta}_{N}}{2} & 0\\
0 & 0 & \frac{\tilde{\delta}_{S}}{\sqrt{2}}
\end{array}
\right)  \equiv\left(
\begin{array}
[c]{ccc}
\delta_{N} & 0 & 0\\
0 & \delta_{N} & 0\\
0 & 0 & \delta_{S}
\end{array}
\right)  \text{  .} \label{eq:expl_sym_br_delta}
\end{align}

Shifting the fields $\sigma_N$ and $\sigma_S$ with their non zero
expectation values $\phi_N$ and $\phi_S$ (\eref{eq:vev}), the
quadratic terms of the Lagrangian, from which the
tree-level meson masses originate, can be determined. The quadratic
terms contain, beside the mixing in the $N-S$ (or $0-8$) sector of the scalar and
pseudoscalar octet, vector-scalar and
axialvector-pseudoscalar mixing terms as well. The later can be
resolved by redefinition of certain (axial-)vector fields (for details
see \cite{elsm_2013}). In our case, only one such field enters in the
calculations, namely the
$a_1^{\mu}$ axialvector meson, which should be redefined as
\be
{a_{1}^{\mu}}^{\pm,0} \longrightarrow {a_{1}^{\mu}}^{\pm,0} + Z_{\pi}
w_{a_{1}} \partial^{\mu} \pi^{\pm,0},\label{eq:a1_shift}
\ee
where
\begin{align}
  Z_{\pi}  &= \frac{m_{a_{1}}}{\sqrt{m_{a_{1}}^{2}-g_{1}^{2}
      \phi_{N}^{2}}},\\
  w_{a_{1}} &= \frac{g_{1}\phi_{N}}{m_{a_{1}}^{2}},
\end{align}
and the $a_1^{\mu}$ axialvector mass is given by
\be
m_{a_{1}}^{2}  = m_{1}^{2} + \frac{1}{2} (2g_{1}^{2} + h_{1} +
h_{2}-h_{3}) \phi_{N}^{2} + \frac{h_{1}}{2} \phi_{S}^{2} + 2\delta_{N}.
\ee
Since in all decuplet decays (\eref{eq:Delta_decay}) a pion is formed,
we also need the explicit expression of the pion mass, which is
\be
m_{\pi}^{2} = Z_{\pi}^{2} \left[m_{0}^{2} + \left(\lambda_{1} +
    \frac{\lambda_{2}}{2} \right) \phi_{N}^{2} + \lambda_{1}
  \phi_{S}^{2} \right].
\ee

The parameters of the meson Lagrangian are determined through the
comparison of the calculated tree-level expressions of the spectrum
and decay widths \cite{elsm_2013} with their experimental value taken
from \cite{PDG}. Some of the parameters are only appear in certain
combinations in every calculated quantities, namely 
\begin{equation}
C_{1} = m_{0}^{2} + \lambda_{1}
\left(\phi_{N}^{2} +\phi_S^2\right)\text{ and } 
C_{2} = m_{1}^{2} + \frac{h_{1}}{2} \left(  \phi_{N}^{2} + \phi_{S}
^{2}\right)
\end{equation}
are such combinations. Moreover without the loss of generality we can
set $\delta_{N}=0$, while all the other meson parameters, taken from
\cite{elsm_2013}, are given in \tabref{Tab:param_meson}

\begin{table}[th]
\centering
\begin{tabular}
[c]{|c|c|}\hline
Parameter & Value\\\hline
$C_{1}$ [GeV$^2$] & $-0.9183 \pm0.0006$ \\\hline
$C_{2}$ [GeV$^2$] &
$0.4135 \pm0.0147$\\\hline
$c_{1}$ [GeV$^{-2}$]& $450.5420 \pm7.0339$\\\hline
$\delta_{S}$ [GeV$^2$] & $0.1511 \pm0.0038$\\\hline
$g_{1}$ & $5.8433 \pm0.0176$\\\hline
$g_{2}$ & $3.0250 \pm0.2329$\\\hline
$\phi_{N}$ [GeV]& $0.1646 \pm0.0001$\\\hline
$\phi_{S}$ [GeV]& $0.1262 \pm0.0001$\\\hline
$h_{2}$ & $9.8796 \pm0.6627$\\\hline
$h_{3}$ & $4.8667 \pm0.0864$\\\hline
$\lambda_{2}$ & $68.2972 \pm0.0435$\\\hline
\end{tabular}
\caption{Meson parameters and their errors}
\label{Tab:param_meson}
\end{table}

\section{On the construction of the Lagrangian}

\label{App:nonlin_expans}

The leading order chiral Lagrangian containing baryon octet, baryon
decuplet and pseudoscalar octet fields is (see e.g. \cite{Semke_thesis})
\begin{eqnarray}
  {\cal{L}}^{(1)}_{\text{chiral}} &=& \tr\left[\bar{B}\left(i\slD - M_{(8)}\right)
    B\right] \no\\
  &-& \tr\left \{ \bar{\Delta}_{\mu}\cdot \left[ \left(i \slD -
        M_{(10)}\right) g^{\mu\nu} - i \left( \gamma^{\mu} D^{\nu} +
        \gamma^{\nu} D^{\mu}\right) + \gamma^{\mu} \left(i \slD +
        M_{(10)} \right) \gamma^{\nu} \right] \Delta_{\nu} \right\} \no\\
  &+& F \tr\left(\bar{B} \gamma^{\mu} \gamma_{5} \left[ iU_{\mu}, B
    \right] \right) +D \tr \left( \bar{B} \gamma^{\mu} \gamma_{5}
    \left \{ iU_{\mu}, B \right\} \right) \label{eq:chiral_Lagr} \\
  &+& C \left\{\tr\left[ \left( \bar{\Delta}_{\mu} \cdot
        i U^{\mu}\right)B\right]+ \text{h.c.} \right\} + H  \tr \left[ \left(
      \bar{\Delta}^{\mu} \cdot \gamma_{\nu}\gamma_{5}
      \Delta_{\mu} \right) i U^{\nu} \right]\no,
\end{eqnarray}
where
\begin{align}
  D_{\mu}B &= \partial_{\mu}B + \Gamma_{\mu} B + B \Gamma_{\mu}^{+},
   \label{eq:DBar}\\
  D_{\mu}\Delta^{ijk}_{\nu} &= \partial_{\mu}\Delta^{ijk}_{\nu} +
  (\Gamma_{\mu})^{i}_{l} \Delta^{ljk}_{\nu} + (\Gamma_{\mu})^{j}_{l}
  \Delta^{ilk}_{\nu} + (\Gamma_{\mu})^{k}_{l}\Delta^{ijl}_{\nu},\label{eq:DDel} \\
  \text{with} &\quad \Gamma_{\mu} = \frac{1}{2}[u^{\dagger}, \partial_{\mu} u] -
  \frac{i}{2}(u^{\dagger} L_{\mu} u + u R_{\mu} u^{\dagger});  \no\\
  U_{\mu} &= -\frac{1}{2}u \left(\nabla_{\mu}U\right)^{\dagger}u,\label{eq:Umu} \\
  \text{with} &\quad \nabla_{\mu}U = \partial_{\mu}U + i \left(UR_{\mu} - L_{\mu}U \right),\no   
\end{align}
and it should be noted that the convention for the left- ($L_{\mu}\equiv
V_{\mu} + A_{\mu}$) and right-handed ($R_{\mu}\equiv V_{\mu} -
A_{\mu}$) fields is just the opposite as in \cite{Semke_thesis}. 
Moreover the $U$ and $u$ fields are defined as
\be
U = e^{i2\tilde{\Phi}/f},\quad u = e^{i\tilde{\Phi}/f}.
\ee
Here $U$ is an $SU(3)$ matrix, which parametrizes the $\tilde{\Phi}$ pseudoscalar
octet non-linearly according to the Callan-Coleman-Wess-Zumino
prescription \cite{Callan_1969}, while the $f$ constant with energy
dimension one is related to the pion decay constant. In order
to get the relevant terms from \eref{eq:chiral_Lagr} it should be
expanded in $\tilde{\Phi}$.
Expanding \eref{eq:DBar}-\eqref{eq:Umu} in $\tilde{\Phi}$ results in
\begin{align}
  D_{\mu}B &= \partial_{\mu} B + i [B,V_{\mu}] + \frac{1}{f}
  \left\{ [A_{\mu},\tilde{\Phi}],B\right\} + \mathcal{O}(\tilde{\Phi}^2),\\
  D_{\mu}\Delta_{\nu}^{ijk} &= \partial_{\mu}\Delta_{\nu}^{ijk} +
  \left(\frac{1}{f}[A_{\mu}, \tilde{\Phi}]^{i}_{l}-iV_{\mu\,l}^{\ \,i}
  \right) \Delta_{\nu}^{ljk} + \left(\frac{1}{f}[A_{\mu}, \tilde{\Phi}]^{j}_{l}-iV_{\mu\,l}^{\ \,j}
  \right) \Delta_{\nu}^{ilk} + \left(\frac{1}{f}[A_{\mu}, \tilde{\Phi}]^{k}_{l}-iV_{\mu\,l}^{\ \,k}
  \right) \Delta_{\nu}^{ijl} + \mathcal{O}(\tilde{\Phi}^2),\\
  U_{\mu} &= \frac{i}{f}\partial_{\mu}\tilde{\Phi} -
  \frac{1}{f}[\tilde{\Phi},V_{\mu}] - i A_{\mu} + \mathcal{O}(\tilde{\Phi}^2),
\end{align}
which should be substituted into \eref{eq:chiral_Lagr} and replace
$\tilde{\Phi}$ by $\Phi$ to get the first three terms of the baryon
Lagrangian \eref{eq:lagr_baryon}.

More details about the chiral Lagrangian and its expansion up to
different orders can be found in \cite{chpt, chptvm, chpt_baryon}.

\section{Two body, tree-level decay width of decuplets}

\label{App:Delta_decay_width}

As can be found in any standard textbook (see
e.g. \cite{textbook:Peskin}) the tree-level two body decay width can be written as
\be
\Gamma_{A\to BC} = I \frac{k}{8\pi m_A^2}\left|\mathcal{M}_{A\to BC}\right|^2,
\ee
where $A$ is the decaying particle, $B$ and $C$ are the resulting
particles $k\equiv k_C = k_B$ is the magnitude of the momentum of the
resulting particles in the restframe of $A$, $\mathcal{M}_{A\to BC}$ is the matrix element and
$I$ is the isospin factor, which shows how many independent decay channels we
have (see later). In our case $A$ is a vectorspinor, $B$ is a
pseudoscalar and $C$ is a spinor field. According to
\eref{eq:lagr_decay_fin}, the structure of the interaction Lagrangian
is $G_{(A\to BC)} A_{\mu}(\partial^{\mu}B) C$, from which the matrix
element can be written as
\be
i\mathcal{M}_{A\to BC} = G_{(A\to BC)}\cdot u_{\mu}^{A}(k_A,s)\cdot
ik_{B}^{\mu}\cdot\bar{u}^C(k_C,s^{\prime}). 
\ee 
Taking the average for the incoming and sum for the outgoing
polarizations the absolute square of the matrix element is given by
\be
\left|\mathcal{M}_{A\to BC}\right|^2 = |G_{(A\to BC)}|^2 \tr\Bigg{\{}
  \frac{1}{4} \underbrace{\sum_{s=-3/2}^{3/2} u_{\mu}^{A}(k_A,s)
  \bar{u}_{\nu}^{A}(k_A, s)}_{-({k \hspace{-0.4em} \slash}_A + m_A)P_{\mu\nu}^A}
\underbrace{ \sum_{s^{\prime} = -1/2}^{1/2} u^{C}(k_C,
  s^{\prime}) \bar{u}^{C}(k_C, s^{\prime})}_{{k \hspace{-0.4em}
    \slash}_C + m_C} \Bigg{\}} k_B^{\mu}k_B^{\nu}, 
\ee
where the $P$ projector is defined as \cite{Projector}
\be
P_{\mu\nu}^A = g_{\mu\nu} - \frac{1}{3} \gamma_{\mu} \gamma_{\nu} -
\frac{2}{3 m_A^2} k_{\mu}^{A} k_{\nu}^{A} + \frac{1}{3 m_A} (
k_{\mu}^{A} \gamma_{\nu} - k_{\nu}^{A} \gamma_{\mu} ).
\ee
After some straightforward calculation the matrix element can be
written as
\begin{align}
  \left|\mathcal{M}_{A\to BC}\right|^2 &= \frac{2}{3} |G_{(A\to BC)}|^2
  k^2 m_A (m_C + E_C),\\
  \text{with,\quad} E_c &= \frac{m_A^2 + m_C^2 - m_B^2}{2m_A}.
\end{align}
Consequently, the decay width reads
\begin{align}
  \Gamma_{A\to BC} &= I \frac{k^3}{12 m_A} (m_C + E_C) |G_{(A\to
    BC)}|^2,\label{eq:form_decay_width}\\
\text{where\quad}  k &= \sqrt{\frac{(m_A^2 - m_B^2 - m_C^2)^2 - 4 m_B^2
    m_C^2}{4 m_A^2}}, \label{eq:k} \\
\text{and  \quad}  E_c &= \frac{m_A^2 + m_C^2 - m_B^2}{2m_A}. \label{eq:E}
\end{align}

\end{document}